\documentclass[MSNbibl,number,citesort,seceqn,dvips]{arxbj}

\aid{0}
\volume{19}
\issue{4}
\pubyear{2013}
\firstpage{1465}
\lastpage{1483}
\doi{10.3150/12-BEJSP10} 

\makeatletter
\newtheorem{theorem}{Theorem}[section]
\newtheorem{proposition}{Proposition}[section]
\newtheorem{lemma}{Lemma}[section]

\newproclaim{definition}{Definition}[section]
\newremark{rem}{Remark}[section]
\newcommand{\eqref}[1]{(\ref{#1})}
\makeatother

\begin{document}
\begin{frontmatter}

\title{Multivariate Bernoulli distribution}
\runtitle{Multivariate Bernoulli distribution}

\begin{aug}
\author[a]{\fnms{Bin}~\snm{Dai}\thanksref{a}\ead[label=e2]{bdai@uwalumni.com}},
\author[b]{\fnms{Shilin} \snm{Ding}\thanksref{b}\ead[label=e3]{dingsl@gmail.com}}
\and
\author[c]{\fnms{Grace} \snm{Wahba}\corref{}\thanksref{c}\ead[label=e1]{wahba@stat.wisc.edu}}
\runauthor{B. Dai, S. Ding and G. Wahba} 
\address[a]{Tower Research Capital, 148 Lafayette Street, FL 12,
New York, NY 10013, USA.\\
\printead{e2}}
\address[b]{Facebook, 1601 Willow Rd,
Menlo Park, CA 94025, USA.
\printead{e3}}
\address[c]{Department of Statistics, University of Wisconsin,
1300 University Ave., Madison, WI 53706, USA.
\printead{e1}}
\end{aug}


\begin{abstract}
In this paper, we consider the multivariate Bernoulli distribution as a
model to estimate the structure of graphs with binary nodes. This
distribution is discussed in the framework of the exponential family,
and its statistical properties regarding independence of the nodes are
demonstrated. Importantly the model can estimate not only the main
effects and pairwise interactions among the nodes but also is capable
of modeling higher order interactions, allowing for the existence of
complex clique effects. We compare the multivariate Bernoulli model
with existing graphical inference models -- the Ising model and the
multivariate Gaussian model, where only the pairwise interactions are
considered. On the other hand, the multivariate Bernoulli distribution
has an interesting property in that independence and uncorrelatedness
of the component random variables are equivalent. Both the marginal and
conditional distributions of a subset of variables in the multivariate
Bernoulli distribution still follow the
multivariate Bernoulli distribution. Furthermore, the multivariate
Bernoulli logistic model is developed under generalized linear model
theory by utilizing the canonical link function in order to include
covariate information on the nodes, edges and cliques. We also consider
variable selection techniques such as LASSO in the logistic model to
impose sparsity structure on the graph. Finally, we discuss extending
the smoothing spline ANOVA approach to the multivariate Bernoulli
logistic model to enable estimation of non-linear effects of the
predictor variables.
\end{abstract}

\begin{keyword}
 \kwd{Bernoulli distribution}
 \kwd{generalized linear models}
 \kwd{LASSO}
 \kwd{smoothing spline}
\end{keyword}

\end{frontmatter}

\section{Introduction} \label{mvb:introduction}
Undirected graphical models have been proved to be useful in a variety
of applications in statistical machine learning. Statisticians and
computer scientists devoted resources to studies in graphs with nodes
representing both continuous and discrete variables. Such models
consider a graph $G = (V, E)$, whose nodes set $V$ represents $K$
random variables $Y_1, Y_2, \ldots, Y_K$ connected or disconnected
defined by the undirected edges set $E$. This formulation allows
pairwise relationships among the nodes to be described in terms of
edges, which in statistics are defined as correlations. The graph
structure can thus be determined under the independence assumptions on
the random variables. Specifically, variables $Y_i$ and $Y_j$ are
conditionally independent given all other variables if the associated
nodes are not linked by an edge. Two important types of graphical
models are the Gaussian model, where the $K$ variables are assumed to
follow a joint multivariate Gaussian distribution, and the Markov model,
which captures the relationships between categorical variables.

However, the assumption that only the pairwise correlations among the
variables are considered may not be sufficient for real applications.
When the joint distribution of the nodes is multivariate Gaussian, the
graph structure can be directly inferred from the inverse of the
covariance matrix of the random variables and in recent years a large
body of literature has emerged in this area for high-dimensional data.
Researchers mainly focus on different sparse structure of the graphs
or, in other words, the covariance matrix for high-dimensional
observations. For example, \cite{Meinshausen:2006} proposes a
consistent approach based on LASSO from \cite{Tibshirani:1996} to
model the sparsity of the graph. Due to the fact that the Gaussian
distribution can be determined by the means and covariance matrix, it
is valid to consider only the pairwise correlations, but this may not
true for some other distributions. The multivariate Bernoulli
distribution discussed in \cite{Whittaker:1990}, which will be studied in
Section~\ref{mvb:formulation}, has a probability density function
involving terms representing third and higher order moments of the
random variables, which is also referred to as clique effects. To
alleviate the complexity of the graph, the so-called Ising model
borrowed from physics gained popularity in the machine learning
literature. \cite{Wainwright:2008} introduces several important
discrete graphical models including the Ising model and \cite
{Banerjee:2008} discussed a framework to infer sparse graph structure
with both Gaussian and binary variables. In this paper, higher than
second interactions among a group of binary random variables are
studied in detail. The multivariate Bernoulli model is equivalent to
Ising model and other undirected graphical model with binary nodes,
which has been used in the machine learning community for various
applications. It can be extended to include $k$-node cliques by adding
monomials of up to $k$ orders \cite{Wainwright:2008}. The Ising model
assumes the nodes
taking values in $\{
-1, 1\}$,
which makes the interpretation of the interactions different form the
multivariate Bernoulli model. The literature related to structure
selection of Ising models and the applications include but are not
limited to \cite{Ravikumar:2010} and \cite{Xue:2012}.

What's more, in some real applications, people are not only interested
in the graph structure but also want to include predictor variables
that potentially have influence on the graph structure. \cite
{Gao:2001} considers a multivariate Bernoulli model which uses a
smoothing spline ANOVA model to replace the linear predictor \cite
{McCullagh:1989} for main effects on the nodes, but set the second and
higher order interactions between the nodes as constants. Higher order
outcomes with hierarchical structure assumptions on the graph involving
predictor variables are studied in \cite{Ding:2011}.

This paper aims at building a unified framework of a generalized linear
model for the multivariate Bernoulli distribution which includes both
higher order interactions among the nodes and covariate information.
The remainder is organized as follows. Section~\ref{mvb:bibernoulli}
starts from the simplest multivariate Bernoulli distribution, the
so-called bivariate Bernoulli distribution, where there are only two
nodes in the graph. The mathematical formulation and statistical
properties of the multivariate Bernoulli distribution are addressed in
Section~\ref{mvb:formulation}. Section~\ref{mvb:comparison} serves to
get a better understanding of the differences and similarities of the
multivariate Bernoulli distribution with the Ising and multivariate
Gaussian models. Section~\ref{mvb:glm} extends the model to include
covariate information on the nodes, edges and cliques, and discusses
parameter estimation, optimization and associated problems in the
resulting multivariate Bernoulli logistic model. Finally,
Section~\ref{mvb:conclusion}
provides conclusion of the paper and some proofs are deferred to \hyperref[app]{Appendix}.

\section{Bivariate Bernoulli distribution}\label{mvb:bibernoulli}
To start from the simplest case, we extend the widely used univariate
Bernoulli distribution to two dimensions in this section and the more
complicated multivariate Bernoulli distribution is explored in Section
\ref{mvb:formulation}. The Bernoulli random variable $Y$, is one with
binary outcomes chosen from $\{0, 1\}$ and its probability density
function is
\[
f_Y(y) = p^y(1-p)^{1-y}.
\]
Next, consider bivariate Bernoulli random vector $(Y_1, Y_2)$, which
takes values from $(0, 0)$, $(0, 1)$, $(1, 0)$ and $(1, 1)$ in the
Cartesian product space $\{0, 1\}^2 = \{0, 1\} \times\{0, 1\}$. Denote
$p_{ij} = P(Y_1 = i, Y_2 = j)$, $i,j = 0,1$, then its probability
density function can be written as
%
\begin{eqnarray}\label{mvb:BBpdf}
P(Y = y) &=& p(y_1, y_2)
\nonumber
\\
&=& p_{11}^{y_1y_2}p_{10}^{y_1(1-y_2)}p_{01}^{(1-y_1)y_2}p_{00}^{(1-y_1)(1-y_2)}
\\
&=& \exp \biggl\{\log(p_{00}) + y_1\log \biggl(
\frac{p_{10}}{p_{00}} \biggr) + y_2\log \biggl(\frac{p_{01}}{p_{00}} \biggr)
+y_1y_2\log \biggl(\frac
{p_{11}p_{00}}{p_{10}p_{01}} \biggr) \biggr\},
\nonumber
\end{eqnarray}
where the side condition $p_{00} + p_{10} + p_{01} + p_{11} = 1$ holds
to ensure it is a valid probability density function.

To simplify the notation, define the natural parameters $f$'s from
general parameters as follows:
%
\begin{eqnarray}
\label{mvb:f1} f^1 &=& \log \biggl(\frac{p_{10}}{p_{00}} \biggr),
\\[5pt]
\label{mvb:f2} f^2 &=& \log \biggl(\frac{p_{01}}{p_{00}} \biggr),
\\[5pt]
\label{mvb:f12}f^{12} &=& \log \biggl(\frac
{p_{11}p_{00}}{p_{10}p_{01}} \biggr),
\end{eqnarray}
and it is not hard to verify the inverse of the above formula
%
\begin{eqnarray}
p_{00} = \frac{1}{1 + \exp(f^1) + \exp(f^2) + \exp(f^1 + f^2 +
f^{12})},
\\
p_{10} = \frac{\exp(f^1)}{1 + \exp(f^1) + \exp(f^2) + \exp(f^1 + f^2 +
f^{12})},
\\
p_{01} = \frac{\exp(f^2)}{1 + \exp(f^1) + \exp(f^2) + \exp(f^1 + f^2 +
f^{12})},
\\
p_{11} = \frac{\exp(f^1 + f^2 + f^{12})}{1 + \exp(f^1) + \exp(f^2) +
\exp(f^1 + f^2 + f^{12})}.
\end{eqnarray}

Here the original density function \eqref{mvb:BBpdf} can be viewed as a
member of the exponential family, and represented in a log-linear
formulation as:
%
\begin{equation}
\label{mvb:BBloglinear}P(Y = y) = \exp \bigl\{\log(p_{00}) +
y_1f^1 + y_2f^2
+y_1y_2 f^{12} \bigr\}.
\end{equation}

Consider the marginal and conditional distribution of $Y_1$ in the
random vector $(Y_1, Y_2)$, we have
%
\begin{proposition}\label{mvb:BBmarginal}
The marginal distribution of $Y_1$ in a bivariate Bernoulli vector
$(Y_1, Y_2)$ following density function \eqref{mvb:BBpdf} is univariate
Bernoulli with density
%
\begin{equation}
\label{mvb:BBmarginalpdf} P(Y_1 = y_1) =
(p_{10} + p_{11})^{y_1}(p_{00} +
p_{01})^{(1-y_1)}.
\end{equation}
What's more, the conditional distribution of $Y_1$ given $Y_2$ is also
univariate Bernoulli with density
%
\begin{eqnarray}
\label{mvb:BBconditionalpdf} P(Y_1 = y_1 |
Y_2 = y_2) = \biggl(\frac{p(1, y_2)}{p(1, y_2) + p(0,
y_2)}
\biggr)^{y_1} \biggl(\frac{p(0, y_2)}{p(1, y_2) + p(0, y_2)} \biggr)^{1-y_1}.
\end{eqnarray}
\end{proposition}

The proposition implies that the bivariate Bernoulli distribution is
similar to the bivariate Gaussian distribution, in that both the
marginal and conditional distributions are still Bernoulli distributed.
On the other hand, it is also important to know under what conditions
the two random variables $Y_1$ and $Y_2$ are independent.

\begin{lemma}\label{mvb:biind}
The components of the bivariate Bernoulli random vector $(Y_1, Y_2)$
are independent if and only if $f^{12}$ in \eqref{mvb:BBloglinear} and
defined in \eqref{mvb:f12} is zero.
\end{lemma}

The Lemma~\ref{mvb:biind} is a special case for Theorem \ref
{mvb:independence} in Section~\ref{mvb:formulation}, and the proof is
attached in \hyperref[app]{Appendix}. It is not hard to see from the
log-linear formulation \eqref{mvb:BBloglinear} that when $f^{12} = 0$,
the probability density function of the bivariate Bernoulli is
separable in $y_1$ and $y_2$ so the lemma holds. In addition, a simple
calculation of covariance between $Y_1$ and $Y_2$ gives
%
\begin{eqnarray}
\label{mvb:covariance} \operatorname{cov}(Y_1, Y_2) &=& E
\bigl[Y_1 - (p_{11}+p_{10})\bigr]
\bigl[Y_2 - (p_{11} + p_{01})\bigr]
\nonumber
\\[-8pt]
\\[-8pt]
\nonumber
&=& p_{11}p_{00} - p_{01}p_{10},
\end{eqnarray}
and using \eqref{mvb:f12}, the disappearance of $f^{12}$ indicates that
the correlation between $Y_1$ and $Y_2$ is null. When dealing with the
multivariate Gaussian distribution, the uncorrelated random variables
are independent as well and Section~\ref{mvb:formulation} below shows
uncorrelatedness and independence is also equivalent for the
multivariate Bernoulli distribution.

The importance of Lemma~\ref{mvb:biind} was explored in \cite
{Whittaker:1990} where it was referred to as Proposition~2.4.1. The
importance of $f^{12}$ (denoted as \textit{u-terms}) is discussed and
called \textit{cross-product ratio} between $Y_1$ and $Y_2$. The same
quantity is actually \textit{log odds} described for the univariate
case in \cite{McCullagh:1989} and for the multivariate case in \cite{Ma:2010}.

\section{Formulation and statistical properties} \label{mvb:formulation}
\subsection{Probability density function}
As discussed in Section~\ref{mvb:bibernoulli}, the two dimensional
Bernoulli distribution possesses good properties analogous to the
Gaussian distribution. This section is to extend it to high-dimensions
and construct the so-called multivariate Bernoulli distribution.

Let $ Y= (Y_1, Y_2, \ldots, Y_K)$ be a $K$-dimensional random vector of
possibly correlated Bernoulli random variables (binary outcomes) and
let $y = (y_1,\ldots
,y_K)$ be a realization of $Y$. The most general form $p(y_1, \ldots,
y_K)$ of the joint probability density is
\begin{eqnarray*}
P(Y_1 = y_1, Y_2 = y_2, \ldots,
Y_K = y_K) &=& p(y_1, y_2,
\ldots, y_K)
\\
&=& p(0, 0, \ldots,0)^{[\prod_{j=1}^K (1-y_j)]}
\\
& &{}\times p(1, 0, \ldots, 0)^{[y_1 \prod_{j=2}^K(1-y_j)]}
\\
& &{}\times p(0, 1, \ldots, 0)^{[(1-y_1)y_2 \prod_{j=3}^K(1-y_j)]}\cdots
\\
& & {}\times p(1, 1, \ldots, 1) ^{[\prod_{j=1}^K y_j]},
\end{eqnarray*}
or in short
%
\begin{eqnarray}
\label{mvb:pdf}p(y) = p_{0, 0, \ldots,0}^{[\prod_{j=1}^K (1-y_j)]} p_{1, 0, \ldots, 0}^{[y_1 \prod_{j=2}^K(1-y_j)]}
p_{0, 1, \ldots, 0}^{[(1-y_1)y_2 \prod_{j=3}^K(1-y_j)]} \cdots p_{1, 1, \ldots, 1} ^{[\prod_{j=1}^K
y_j]}.
\end{eqnarray}

To simplify the notation, denote the quantity $S$ to be
%
\begin{equation}
\label{mvb:bigS} S^{j_1j_2\cdots j_r} = \sum_{1\le s\le r}
f^{j_s} + \sum_{1\le
s<t\le r} f^{j_sj_t} +
\cdots+ f^{j_1j_2\cdots j_r},
\end{equation}
and in the bivariate Bernoulli case $S^{12} = f^1 + f^2 + f^{12}$. To
eliminate the product in the tedious exponent of \eqref{mvb:pdf},
define the interaction function $B$
%
\begin{equation}
\label{mvb:bigB} B^{j_1j_2\cdots j_r}(y) = y_{j_1}y_{j_2}\cdots
y_{j_r},
\end{equation}
so correspondingly in the bivariate Bernoulli distribution for the
realization $(y_1, y_2)$ of random vector $(Y_1, Y_2)$, the interaction
function of order 2 is $B^{12}(y) = y_1y_2$.\vspace*{1pt} This is the only available
order two interaction for the bivariate case. In general, there are
$\bigl({K\atop 2}\bigr) = \frac{K(K-1)}{2}$ different second interactions among
the binary components of the multivariate Bernoulli random vector.

The log-linear formulation of the multivariate Bernoulli distribution
induced from \eqref{mvb:pdf} is
%
\begin{eqnarray}
\label{mvb:MBloglinear} l(y,\mathbf{f}) &=& -\log\bigl[p(y)\bigr]
\nonumber
\\[-8pt]
\\[-8pt]
\nonumber
\label{eq2} &=& - \Biggl[\sum_{r=1}^K
\biggl(\sum_{1\leq j_1<j_2<\cdots
<j_r\leq K}f^{j_1j_2\cdots j_r}B^{j_1j_2\cdots
j_r}(y)
\biggr)- b({\mathbf{f}}) \Biggr],
\end{eqnarray}
where ${\mathbf{f}} = (f^1,f^2,\ldots, f^{12\cdots K})^T$ is the vector
of the natural parameters for multivariate Bernoulli, and the
normalizing factor $b({\mathbf{f}})$ is defined as
%
\begin{equation}
\label{mvb:b}b({\mathbf{f}}) = \log\sum_{r=1}^K
\biggl[1+ \biggl(\sum_{1\leq j_1<j_2<\cdots<j_r\leq K}\exp\bigl[S^{j_1j_2\cdots j_r}
\bigr] \biggr) \biggr].
\end{equation}

As a member of the exponential distribution family, the multivariate
Bernoulli distribution has the fundamental `link' between the natural
and general parameters.

\begin{lemma}[(Parameter transformation)] \label{mvb:transform}
For the multivariate Bernoulli model, the
general parameters and natural parameters have the following relationship.
\begin{eqnarray*}\label{mvb:f}
&&\hspace*{-4pt}\exp\bigl( f^{j_1j_2\cdots j_r}\bigr)
\nonumber
\\[-8pt]
\\[-8pt]
\nonumber
&&\hspace*{-4pt}\quad =\frac{\prod p(\mbox{even \# zeros among~}j_1,j_2,\ldots,j_r
\mbox{ components and other components are all zero})}{\prod p(\mbox{odd \# zeros among~}j_1,j_2,\ldots,j_r \mbox{~components and
other components are all zero})},
\end{eqnarray*}
where \# refers to the number of zeros among the superscript
$y_{j_1}\cdots y_{j_r}$ of $f$. In addition,
%
\begin{eqnarray}
\label{mvb:S} &&\exp\bigl(S^{j_1j_2\cdots j_r}\bigr)
\nonumber
\\[-8pt]
\\[-8pt]
\nonumber
&&\qquad = \frac{ p(j_1,j_2,\ldots,j_r \mbox{ positions are
one, others are zero})}{p(0,0,\ldots,0)}
\end{eqnarray}
and conversely the general parameters can be represented by the natural
parameters
\begin{eqnarray}
\label{mvb:p}&& p(j_1,j_2,\ldots,j_r
\mbox{ positions are one, others are zero})
\nonumber
\\[-8pt]
\\[-8pt]
\nonumber
&&\qquad= \frac{\exp(S^{j_1j_2\cdots j_r})}{\exp (b({\mathbf{f}}) )}.
\end{eqnarray}
\end{lemma}

Based on the log-linear formulation \eqref{mvb:MBloglinear} and the
fact that the multivariate Bernoulli distribution is a member of the
exponential family, the interactions functions $B^{j_1j_2\cdots
j_r}(y)$ for all combinations $j_1j_2\cdots j_r$ define the sufficient
statistics. In addition, the log-partition function $b({\mathbf{f}})$
as in~\eqref{mvb:b} is useful to determine the expectation and variance
of the sufficient statistics to be addressed in later sections.

\subsection{Independence, marginal and conditional distributions}
One of the most important statistical properties for the multivariate
Gaussian distribution is the equivalence of independence and
uncorrelatedness. As a natural multivariate extension of the univariate
Bernoulli distribution, it is of great interest to explore independence
among components of the multivariate Bernoulli distribution and it is
the topic for this section.

The independence of components of a random vector is determined by
separability of coordinates in its probability density function and it
is hard to get directly from \eqref{mvb:pdf}. However, based on the
relationship between the natural parameters and the outcome in the
log-linear formulation~\eqref{mvb:MBloglinear}, the independence
theorem of the distribution can be derived as follows with proof
deferred to \hyperref[app]{Appendix}.
%
\begin{theorem}[(Independence of Bernoulli outcomes)]\label{mvb:independence}
For the multivariate Bernoulli
distribution, the random vector $ Y= (Y_1, \ldots, Y_K)$ is independent
element-wise if and only if
%
\begin{equation}
\label{mvb:independence1} f^{j_1j_2\cdots j_r} = 0\qquad \forall 1\leq
j_1<j_2<\cdots<j_r\leq K, r \geq2.
\end{equation}
In addition, the condition in equation \eqref{mvb:independence1} can be
equivalently written as
%
\begin{equation}
\label{mvb:independence2} S^{j_1j_2\cdots j_r} = \sum_{k=1}^r
f^{j_k} \qquad\forall r \geq2.
\end{equation}
%
\end{theorem}

The importance of the theorem is to link the independence of components
of a random vector following the multivariate Bernoulli distribution to
the natural parameters. Notice that to ensure all the single random
variable to be independent of all the others is a strong assertion and
in graphical models, researchers are more interested in the
independence of two groups of nodes, so we have the following theorem:
%
\begin{theorem}[(Independence of groups)]\label{mvb:independence_group}
For random vector $ Y= (Y_1, \ldots, Y_K)$
following the multivariate Bernoulli distribution, without of loss of
generality, suppose two blocks of nodes $Y' = (Y_1, Y_2, \ldots, Y_r)$,
$Y'' = (Y_{r+1}, Y_{r+2}, \ldots, Y_s)$ with $1\leq r < s \leq K$, and
denote index set $\tau_1 = \{1, 2, \ldots, r\}$ and $\tau_2 = \{r+1,
r+2, \ldots, s\}$. Then $Y'$ and $Y''$ are independent if and only if
%
\begin{equation}
\label{mvb:ind_group} f^\tau= 0\qquad \forall \tau\cap
\tau_1\neq\emptyset \mbox{ and } \tau\cap\tau_2\neq\emptyset.
\end{equation}
\end{theorem}

The proof of Theorem~\ref{mvb:independence_group} is also deferred to
\hyperref[app]{Appendix}. The theorem delivers the message that the two groups
of binary nodes in a graph are independent if all the natural
parameters $f$'s corresponding to the index sets that include indices
from both groups disappear.

Furthermore, analogous to the multivariate Gaussian distribution,
researchers are interested in statistical distributions of marginal and
conditional distributions for the multivariate Bernoulli distribution.
Likewise, the multivariate Bernoulli distribution maintains the good
property that both the marginal and conditional distributions are still
multivariate Bernoulli as stated in the following proposition.

\begin{proposition}\label{mvb:marginal}
The marginal distribution of the random vector $(Y_1, \ldots, Y_K)$
which follows multivariate Bernoulli distribution with density function
\eqref{mvb:pdf} to any order is still a \textup{multivariate Bernoulli}
with density
%
\begin{equation}
P(Y_1 = y_1, Y_2 = y_2, \ldots,
Y_r = y_r) = \sum_{y_{r+1}}\cdots
\sum_{y_K} p(y_1, \ldots, y_K)
\end{equation}
for some $r < K$.

What's more, the conditional distribution of $(Y_1, Y_2, \ldots, Y_r)$
given the rest is also \textup{multivariate Bernoulli} with density
%
\begin{equation}
P(Y_1 = y_1 ,\ldots, Y_r = y_r |
Y_{r+1} = y_{r+1}, \ldots, Y_K = y_K) =
\frac{p(y_1, \ldots, y_K)}{p(y_{r+1}, \ldots, y_K)}.
\end{equation}
\end{proposition}

\subsection{Moment generating functions}
The moment generating function for the multivariate Bernoulli
distribution is useful when dealing with moments and proof of Theorem
\ref{mvb:independence}.
%
\begin{eqnarray}\label{mvb:mgf}
\psi(\mu_1, \mu_2, \ldots, \mu_K) &=& E
\bigl[\exp (\mu_1Y_1 + \mu _2Y_2 +
\cdots+ \mu_KY_K ) \bigr]
\nonumber
\\
&=& p_{00\cdots0}e^0 + p_{10\cdots0}e^{\mu_1} + \cdots+
p_{11\cdots
1}e^{\mu_1+ \mu_2+\cdots+\mu_K}
\\
 &=& \sum_{r=1}^K\sum
_{j_1\le j_2\le\cdots\le j_r} \frac
{\exp[S^{j_1j_2\cdots j_r}]}{\exp[b(\mathbf{f})]}\exp \Biggl[\sum
_{k=1}^r\mu_{j_k} \Biggr].\nonumber
\end{eqnarray}
Hence, from the formula the moment generating function is solely
determined by the $S$ functions, which are the transformation of the
natural parameters $f$'s.

\subsection{Gradient and Hessian}
As a member of the exponential family, the gradient and Hessian (Fisher
information) are the mean and covariance of the random vector $(Y_1,
Y_2, \ldots, Y_K)$. Therefore, they are important in statistics but
also crucial for model inference when the proper optimization problem
is established. To examine the formulation of gradient and Hessian for
the logarithm of the multivariate Bernoulli distribution \eqref
{mvb:pdf}, let us define some notations.

Denote $\mathcal{T}$ to be the set of all possible superscripts of the
$f$'s including the null superscript with $f^\emptyset= 0$, so it has
$2^K$ elements. In other words, $\mathcal{T}$ is the power set of
indices $\{1, 2, \ldots, K\}$. Let $|\cdot|$ be the cardinality of a
set then $|\mathcal{T}| = 2^K$. We can define the relation subset
$\subset$ for $\tau_1,\tau_2\in\mathcal T$ as follows.

\begin{definition}\label{mvb:subset}
For any two superscripts $\tau_1 = \{j_1,j_2,\ldots, j_r\}$ such that
$\tau_1\in\mathcal T$ and $\tau_2 = \{k_1,k_2,\ldots, k_s\}$ with $\tau
_2\in\mathcal T$ and $r \leq s$, we say that $\tau_1\subseteq\tau_2$ if for
any $j\in\tau_1$, there is a $k\in\tau_2$ such that $j=k$.
\end{definition}

Based on the definition, the $S$'s in \eqref{mvb:bigS} can be
reformulated as
%
\begin{equation}
\label{mvb:simpleS} S^\tau= \sum_{\tau_0\subseteq\tau}f^{\tau_0},
\end{equation}
specifically, $S^\emptyset= 0$. Consider the gradient of the
log-linear form \eqref{mvb:MBloglinear} with respect to the $f$'s, for
any $\tau\in\mathcal T$,
%
\begin{eqnarray}\label{mvb:gradient}
\frac{\partial l(y,\mathbf{f})}{\partial f^\tau} &=& -B^\tau(y) + \frac
{\partial b(\mathbf{f})}{\partial f^\tau}
\nonumber
\\[-8pt]
\\[-8pt]
\nonumber
 &=& -B^\tau(y) + \frac{\sum_{\tau_0\supseteq\tau
}{\exp[S^{\tau_0}]}}{b({\mathbf{f}})}.
\end{eqnarray}

The derivation of partial derivative of $b$ with respect to $f^\tau$ in
\eqref{mvb:gradient} is
%
\begin{eqnarray}
\frac{\partial b(\mathbf{f})}{\partial f^\tau} &=& \frac{1}{\exp
[b(\mathbf{f})]} \cdot\frac{\partial\exp[b(\mathbf{f})]}{\partial
f^\tau}
\nonumber
\\
&=& \frac{1}{\exp[b(\mathbf{f})]} \cdot\frac{\partial \sum_{\tau_0\in
\mathcal{T}}\exp[S^{\tau_0}]}{\partial f^\tau}
\nonumber
\\[-8pt]
\\[-8pt]
\nonumber
&=& \frac{\sum_{\tau_0\supseteq\tau}{\exp[S^{\tau_0}]}}{\exp[b(\mathbf
{f})]}
\\
&=& E\bigl[B^\tau(y)\bigr],
\nonumber
\end{eqnarray}
and the result can also be derived from the moment generating function
\eqref{mvb:mgf} by taking derivatives with respect to the $\mu$'s.

A simple example of \eqref{mvb:gradient} in the bivariate Bernoulli
distribution \eqref{mvb:BBloglinear} is
\[
\frac{\partial l(y,\mathbf{f})}{\partial f^1} = -y_1 + \frac{\exp(f^1)
+ \exp(S^{12})}{b({\mathbf{f}})}.
\]

Further, the general formula for the second order derivative of \eqref
{mvb:MBloglinear} with respect to any two natural parameters $f^{\tau
_1}$ and $f^{\tau_2}$ is
%
\begin{eqnarray}\label{mvn:hessian}
\frac{\partial^2 l(y, f)}{\partial f^{\tau_1}\partial f^{\tau_2}} &=& \frac{\partial^2 b(\mathbf{f})}{\partial f^{\tau_1}\partial f^{\tau_2}}
\nonumber
\\
&=& \frac{\partial}{\partial f^{\tau_1}} \biggl(\frac{\sum_{\tau
_0\supseteq\tau_2}{\exp[S^{\tau_0}]}}{\exp[b(\mathbf{f})]} \biggr)
\nonumber
\\[-8pt]
\\[-8pt]
\nonumber
&=& \frac{\sum_{\tau_0\supseteq\tau_1,  \tau_0\supseteq\tau_2}\exp
[S^{\tau_0}]\exp[b(\mathbf{f})] - \sum_{\tau_0\supseteq\tau_1}{\exp
[S^{\tau_0}]}\sum_{\tau_0\supseteq\tau_2}{\exp[S^{\tau_0}]}}{\exp
[2b(\mathbf{f})]}
\\
&=& \operatorname{cov} \bigl(B^{\tau_1}(y),
B^{\tau
_2}(y) \bigr).\nonumber
\end{eqnarray}

In the bivariate Bernoulli distribution,
\[
\frac{\partial^2 l(y, f)}{\partial f^1\partial f^2} = \frac{\exp
[S^{12}]\exp[b(\mathbf{f})] -
(\exp[f^1] + \exp[S^{12}])(\exp[f^2] +
\exp[S^{12}])}{\exp[2b(\mathbf{f})]}.
\]

\section{The Ising and the multivariate Gaussian models} \label{mvb:comparison}

As mentioned in Section~\ref{mvb:introduction}, the Ising and the
multivariate Gaussian distributions are two main tools to study
undirected graphical models, and this section is to compare the
multivariate Bernoulli model introduced in Section \ref
{mvb:formulation} with these two popular models.

\subsection{The Ising model}

The Ising model, which originated from \cite{Ising:1925}, becomes
popular when the graph structure is of interest with nodes taking
binary values. The log-linear density of the random vector $(Y_1, \ldots
, Y_K)$ is
%
\begin{equation}
\label{mvb:ising}\log\bigl[f(Y_1, \ldots, Y_K)\bigr] =
\sum_{j=1}^K\theta _{j,j}Y_j
+ \sum_{1\leq j < j' \leq K} \theta_{j, j'}Y_jY_{j'}
-\log \bigl[Z(\Theta)\bigr],
\end{equation}
where $\Theta= (\theta_{j,j'})_{K\times K}$ is a symmetric matrix
specifying the network structure, but it is not necessarily positive
semi-definite. The log-partition function $Z(\Theta)$ is defined as
%
\begin{equation}
\label{mvb:bigZ} Z(\Theta) = \sum_{Y_j\in\{0,1\}, 1\leq j \leq K}\exp \Biggl(
\sum_{j=1}^K\theta_{j,j}Y_j
+ \sum_{1\leq j<j'\leq K}\theta _{j,j'}Y_jY_{j'}
\Biggr),
\end{equation}
and notice that it is not related to $Y_j$ due to the summation over
all possible values of $Y_j$ for $j = 1, 2, \ldots, K$.

It is not hard to see that the multivariate Bernoulli is an extension
of the Ising model, which assumes all $S^\tau= 0$ for any $\tau$ such
that $|\tau| > 2$ and $\theta_{j,j'} = S^{jj'}$. In other words, in the
Ising model, only pairwise interactions are considered. \cite
{Ravikumar:2010} pointed out that the higher order interactions, which
is referred to as clique effects in this paper, can be converted to
pairwise ones through the introduction of additional variables and thus
retain the Markovian structure of the network defined in \cite
{Wainwright:2008}.

\subsection{Multivariate Gaussian model}
When continuous nodes are considered in a graphical model, the
multivariate Gaussian distribution is important since, similar to the
Ising model, it only considers interactions up to order two. The
log-linear formulation is
%
\begin{equation}
\label{mvb:gaussian} \log\bigl[f(Y_1, \ldots, Y_K)\bigr]
= \bigl(-\tfrac
{1}{2}(Y-\mu)^T\Sigma(Y-\mu) \bigr) - \log
\bigl[Z(\Sigma)\bigr],
\end{equation}
where $Z(\Sigma)$ is the normalizing factor which only depends on the
covariance matrix $\Sigma$.

\subsection{Comparison of different graphical models}
The multivariate Bernoulli \eqref{mvb:MBloglinear}, Ising \eqref
{mvb:ising} and multivariate Gaussian \eqref{mvb:gaussian} are three
different kinds of graphical models and they share many similarities
\begin{enumerate}
\item All of them are members of the exponential family.
\item Uncorrelatedness and independence are equivalent.
\item Conditional and marginal distributions maintain the same structure.
\end{enumerate}

However, some differences do exist. the multivariate Bernoulli and the
Ising models both serve as tools to model graph with binary nodes, and
are certainly different from the multivariate Gaussian model which
formulates continuous variables. In addition, the multivariate
Bernoulli specifies clique effects among nodes whereas the Ising model
simplifies to deal with only pairwise interactions and the multivariate
Gaussian essentially is uniquely determined by its mean and covariance
structure, which is also based on first and second order moments. Table
\ref{mvb:compare} illustrates the number of parameters needed to
uniquely determine the distribution for these models as the number of
nodes $K$ in the graph increases.

\begin{table}[b]
\caption{The number of parameters in the multivariate Bernoulli, the
Ising and the multivariate Gaussian models}
\label{mvb:compare}
\begin{tabular*}{\textwidth}{@{\extracolsep{\fill}}lccc@{}}
\hline
Graph dimension & Multivariate Bernoulli & Ising & Multivariate
Gaussian\\
\hline
$1$ & $1$ & $1$ & $2$ \\
$2$ & $3$ & $3$ & $5$ \\
$3$ & $7$ & $6$ & $9$ \\
$\cdots$ & $\cdots$ & $\cdots$ & $\cdots$\\
$K$ & $2^K - 1$ & $\frac{K(K+1)}{2}$ & $K + \frac{K(K+1)}{2}$\\
\hline
\end{tabular*}
\end{table}

\section{Multivariate Bernoulli logistic models} \label{mvb:glm}
\subsection{Generalized linear model}
As discussed in Section~\ref{mvb:formulation}, the multivariate
Bernoulli distribution is a member of the exponential family and as a
result, the generalized linear model theory in \cite{McCullagh:1989}
applies. The natural parameters ($f$'s) in Lemma~\ref{mvb:transform}
can be formulated as a linear predictor in~\cite{McCullagh:1989} such
that for any $\tau\in\mathcal{T}$ with $\mathcal{T}=\{1, 2, \ldots, K\}$
%
\begin{equation}
\label{mvb:linear} f^{\tau}(x) = c_0^\tau+
c_1^\tau x_1 + \cdots+ c_p^\tau
x_p,
\end{equation}
where the vector $c^\tau= (c_0^\tau, \ldots, c_p^\tau)$ for $\tau\in
\mathcal{T}$ is the coefficient vector to be estimated and $x = (x_1,
x_2, \ldots, x_p)$ is the observed covariate. Here $p$ is the number of
variables and there are $2^K - 1$ coefficient vectors to be estimated
so in total $p \times(2^K - 1)$ unknown parameters. Equation~\eqref{mvb:linear}
is built on the canonical link where natural parameters are directly
modeled as linear predictors, but other links are possible and valid as well.

When there are $n$ samples observed from a real data set with outcomes
denoted as $y(i) = (y_1(i), \ldots, y_K(i))$ and predictor variables
$x(i) = (x_1(i), \ldots, x_p(i))$, the negative log likelihood for the
generalized linear model of the multivariate Bernoulli distribution is
%
\begin{equation}
\label{mvb:linearlike} l\bigl(y,\mathbf{f}(x)\bigr) = \sum
_{i=1}^n \biggl[-\sum_{\tau\in\mathcal{T}}f^\tau
\bigl(x(i)\bigr)B^\tau\bigl(y(i)\bigr) + b\bigl(\mathbf{f}(x)\bigr)
\biggr],
\end{equation}
where, similar to \eqref{mvb:b} the log partition function $b$ is
\[
b\bigl(\mathbf{f}(x)\bigr) = \log \biggl[1 + \sum_{\tau\in\mathcal{T}}
\exp\bigl[S^\tau \bigl(x(i)\bigr)\bigr] \biggr].
\]

When dealing with the univariate Bernoulli distribution using formula
\eqref{mvb:linearlike}, the resulting generalized linear model
corresponding to the multivariate Bernoulli model is the same for
logistic regression. Thus the model is referred to as the multivariate
Bernoulli logistic model in this paper.

\subsection{Gradient and Hessian}
To optimize the negative log likelihood function \eqref{mvb:linear}
with respect to the coefficient vector~$c^\tau$, the efficient and
popular iterative re-weighted least squares algorithm mentioned
in~\cite{McCullagh:1989} can be implemented. Nevertheless, the gradient
vector and Hessian matrix (Fisher Information) with respect to the
coefficients $c^\tau$ are still required.

Consider any\vspace*{1pt} $\tau\in\mathcal{T}$, the first derivative with respect to
$c_j^\tau$ in the negative log likelihood \eqref{mvb:linearlike} of the
multivariate Bernoulli logistic model, according to \eqref
{mvb:gradient} and ignoring index~$i$, is
%
\begin{eqnarray}\label{mvb:linear_gradient}
\frac{\partial l(y,f)}{\partial c^\tau_j} = \frac{\partial
l(y,f)}{\partial f^\tau} \frac{\partial f^\tau}{\partial
c^\tau_j}
= \sum_{i = 1}^n
\biggl[-B^\tau(y) + \frac
{\sum_{\tau_0\supseteq\tau}{\exp[S^{\tau_0}(x)]}}{\exp[b(\mathbf
{f}(x))]} \biggr] x_j.
\end{eqnarray}
Further, the second derivative for any two coefficients $c_j^{\tau_1}$
and $c_k^{\tau_2}$ is
%
\begin{eqnarray}\label{mvb:linear_hessian}
\frac{\partial^2l(y, f)}{\partial c^{\tau_1}_j\partial c^{\tau_2}_k} &=& \frac{\partial}{\partial c^{\tau_1}_j} \biggl(\frac{\partial l(y,
f)}{\partial f^{\tau_2}}
\frac{\partial f^{\tau_2}}{\partial c^{\tau
_2}_k} \biggr)
\nonumber
\\
&=& \frac{\partial f^{\tau_1}}{\partial c^{\tau_1}_j}\frac{\partial^2
l(y, f)}{\partial f^{\tau_1}\partial f^{\tau_2}}\frac{\partial
f^{\tau_2}}{\partial c^{\tau_2}_k}
\nonumber
\\[-8pt]
\\[-8pt]
\nonumber
&=& \sum_{i=1}^n \frac{\partial^2 l(y, f)}{\partial f^{\tau_1}\partial
f^{\tau_2}}
x_j x_k
\\
&=& \frac{\sum_{\tau_0\supseteq\tau_1,
\tau_0\supseteq\tau_2}\exp[S^{\tau_0}(x)]}{\exp[b(f(x))]} x_j x_k -
\frac{\sum_{\tau_0\supseteq\tau_1}{\exp
[S^{\tau_0}(x)]}\sum_{\tau_0\supseteq\tau_2}{\exp[S^{\tau_0}(x)]}}{\exp
[2b(f(x))]} x_j
x_k.\nonumber
\end{eqnarray}

\subsection{Parameters estimation and optimization} \label{mvb:estimation}
With gradient \eqref{mvb:linear_gradient} and Hessian \eqref
{mvb:linear_hessian} at hand, the minimization of the negative log
likelihood~\eqref{mvb:linearlike} with respect to the coefficients
$c^\tau$ can be solved with Newton--Raphson or the Fisher's scoring
algorithm (iterative re-weighted least squares) when the Hessian is
replaced by the Fisher information matrix. Therefore, in every
iteration, the new step size for current estimate $\hat c^{(s)}$ is
computed as
%
\begin{equation}
\label{mvb:step} \triangle c = - \biggl(\frac{\partial^2l(y,
f)}{\partial c^{\tau_1}_j\partial c^{\tau_2}_k} \bigg|_{c=\hat
c^{(s)}}
\biggr)^{-1} \cdot \biggl(\frac{\partial l(y,f)}{\partial c^\tau
_j} \bigg|_{c=\hat c^{(s)}} \biggr).
\end{equation}
The process continues until the convergence criterion is met.

\subsection{Variable selection}
Variable selection is important in modern statistical inference. It is
also crucial to select only the significant variables to determine the
structure of the graph for better model identification and prediction
accuracy. The pioneering paper \cite{Tibshirani:1996} introduced the
LASSO approach to linear models. Various properties of the method were
demonstrated such as in \cite{Zhao:2006} and extensions of the model
to different frameworks were discussed in \cite{Meinshausen:2006,Zhao:2007,Park:2008} etc.

The approach can be extended to the multivariate Bernoulli distribution
since it is a member of the exponential family. What we have to do is
to apply the $l_1$ penalty to the coefficients in \eqref{mvb:linear},
and the target function is
%
\begin{equation}
\label{mvb:lasso} L_\lambda(x, y) = \frac{1}{n}\sum
_{i=1}^nl\bigl(y(i),\mathbf{f}\bigl(x(i)\bigr)\bigr)
+ \sum_{\tau\in\mathcal{T}}\lambda_\tau\sum
_{j=1}^p\bigl|c_j^\tau\bigr|,
\end{equation}
where $\lambda_\tau$ are the tuning parameters need to be chosen
adaptively. The superscript $\tau$ allows flexibility to have
natural\vadjust{\goodbreak}
parameters with different levels of complexity. For tuning in penalized
regression problems, the randomized generalized approximate
cross-validation (GACV) designed for smoothing spline models introduced
in \cite{Xiang:1994} can be derived for LASSO problem, such as in
\cite{Shi:2008}. The widely used information criterion AIC and BIC can
also be implemented, but the degrees of freedom cannot be calculated
exactly. \cite{Ma:2010} demonstrates that the number of nonzero
estimates can serve as a good approximation in the multivariate
Bernoulli logistic model. There are several efficient algorithms
proposed to optimize the problem \eqref{mvb:lasso}, for example, the
LASSO-pattern search introduced in \cite{Shi:2008} can handle large
number of unknowns provided that it is known that at most a modest
number are nonzeros. Recently, \cite{Shi:2012} has extended the
algorithm in
\cite{Shi:2008} to the scale of multi-millions of unknowns. Coordinate
descent \cite{Friedman:2010} is also proven to be fast in solving
large $p$ small $n$ problems.

\subsection{Smoothing spline ANOVA model}
The smoothing spline model gained popularity in non-linear statistical
inference since it was proposed in \cite{Craven:1978} for univariate
predictor variables. More importantly, multiple smoothing spline models
for generalized linear models enable researchers to study complex real
world data sets with increasingly powerful computers as described in
\cite{Wahba:1995}.

As a member of the exponential family, the multivariate Bernoulli
distribution can be formulated under smoothing spline ANOVA framework.
\cite{Gao:2001} considers the smoothing spline ANOVA multivariate
Bernoulli model but the interactions are restricted to be constant.
However, in general the natural parameters or linear predictors $f$'s
can be relaxed to reside in a reproducing kernel Hilbert space. That is
to say, for the observed predictor vector $x$, we have
%
\begin{equation}
\label{mvb:splineANOVA} f^\tau(x) = \eta^\tau(x)\qquad \mbox{with }
\eta^\tau\in\mathcal{H}^\tau , \tau\in\mathcal{T},
\end{equation}
where $\mathcal{H}^\tau$ is a reproducing kernel Hilbert space and the
superscript $\tau$ allows a more flexible model such that the natural
parameters can come from different reproducing kernel Hilbert spaces.
Further, $\mathcal{H}^\tau$ can be formulated to have several
components to handle multivariate predictor variables, that is $\mathcal
{H}^\tau= \oplus_{\beta=0}^p\mathcal{H}^\tau_\beta$ and details can be
found in \cite{Gu:2002}.

As a result, the $\eta^\tau$ is estimated from the variational problem
%
\begin{equation}
\label{mvb:splineTarget} \mathcal{I_\lambda}(x, y) = \frac{1}{n}
\sum_{i=1}^nl\bigl(y(i),\bolds{\eta }
\bigl(x(i)\bigr)\bigr) + \lambda J(\bolds{\eta}),
\end{equation}
where $\bolds{\eta}$ is the vector form of $\eta^\tau$'s. The penalty
is seen to be
%
\begin{eqnarray}
\label{mvb:penalty} \lambda J(\bolds{\eta}) = \lambda\sum
_{\tau\in\mathcal{T}}\theta_\tau ^{-1}\bigl\Vert P_1^\tau
\eta^\tau\bigr\Vert^2
\end{eqnarray}
with $\lambda$ and $\theta_\tau$ being the smoothing parameters. This
is an over-parameterization adopted in \cite{Gu:2002}, as what really
matters are the ratios $\lambda/\theta_\tau$. The functional $P_1^\tau$
projects function $\eta^\tau$ in $\mathcal{H}^\tau$ onto the smoothing
subspace $\mathcal{H}_1^\tau$.

By the argument of smoothing spline ANOVA model in \cite{Gu:2002}, the
minimizer $\eta^\tau$ has the expression as in \cite{Wahba:1990},
%
\begin{equation}
\label{mvb:lineareta} \eta^\tau(x) = \sum_{\nu= 1}^md_\nu^\tau
\phi_\nu^\tau(x) + \sum_{i=1}^nc_i^\tau
R^\tau(x_i, x),
\end{equation}
where $\{\phi^\tau_\nu\}_{\nu=1}^m$ is a basis of $\mathcal{H}_0^\tau=
\mathcal{H}^\tau\ominus\mathcal{H}^\tau_1$, the null space
corresponding to the projection functional $P_1^\tau$. $R^\tau(\cdot,
\cdot)$ is the reproducing kernel for $\mathcal{H}^\tau_1$.

The variational problem \eqref{mvb:splineTarget} utilizing the
smoothing spline ANOVA framework can be solved by iterative re-weighted
least squares \eqref{mvb:step} due to the linear formulation \eqref
{mvb:lineareta}. More on tuning and computations including software
will appear in \cite{Dai:2012}.

\section{Conclusion} \label{mvb:conclusion}
We have shown that the multivariate Bernoulli distribution, as a member
of the exponential family, is a way to formulate the graph structure of
binary variables. It can not only model the main effects and pairwise
interactions as the Ising model does, but also is capable of estimating
higher order interactions. Importantly, the independence structure of
the graph can be modeled via significance of the natural parameters.
The most interesting observation of the multivariate Bernoulli
distribution is its similarity to the multivariate Gaussian
distribution. Both of them have the property that independence and
uncorrelatedness of the random variables are equivalent, which is
generally not true for other distributions. In addition, the marginal
and conditional distributions of a subset of variables still follow the
multivariate Bernoulli distribution.

Furthermore, the multivariate Bernoulli logistic model extends the
distribution to a generalized linear model framework to include effects
of predictor variables. Under this model, the traditional statistical
inferences such as point estimation, hypothesis test and confidence
intervals can be implemented as discussed in \cite{McCullagh:1989}.

Finally, we consider two extensions to the multivariate Bernoulli
logistic model. First, the variable selection technique using LASSO can
be incorporated to enable finding important patterns from a large
number of candidate covariates. Secondly, the smoothing spline ANOVA
model is introduced to consider non-linear effects of the predictor
variables in nodes, edges and cliques level.

\begin{appendix}\label{app}
\section*{Appendix: Proofs}\label{appendix:proof}
\begin{pf*}{Proof of Proposition~\ref{mvb:BBmarginal}}
With the joint density function of the random vector $(Y_1, Y_2)$, the
marginal distribution of $Y_1$ can be derived
\begin{eqnarray*}
P(Y_1 = 1) &=& P(Y_1 = 1, Y_2 = 0) +
P(Y_1 = 1, Y_2 = 1)
\\
&=& p_{10} + p_{11}.
\end{eqnarray*}
Similarly,
\[
P(Y_1 = 0) = p_{00} + p_{11}.
\]
Combining the side condition of the parameters $p$'s,
\[
P(Y_1 = 1) + P(Y_1 = 0) = p_{00} +
p_{01} + p_{10} + p_{11} = 1.
\]
This demonstrates that $Y_1$ follows the univariate Bernoulli
distribution and its density function is~\eqref{mvb:BBmarginal}.

Regarding the conditional distribution, notice that
\begin{eqnarray*}
P(Y_1 = 0 | Y_2 = 0) &=& \frac{P(Y_1 = 0, Y_2 = 0)}{P(Y_2 = 0)}
\\
&=& \frac{p_{00}}{p_{00} + p_{10}},
\end{eqnarray*}
and the same process can be repeated to get
\[
P(Y_1 = 1 | Y_2 = 0) = \frac{p_{10}}{p_{00} + p_{10}}.
\]
Hence, it is clear that with condition $Y_2 = 0$, $Y_1$ follows a
univariate Bernoulli distribution as well. The same scenario can be
examined for the condition $Y_2 = 1$. Thus, the conditional
distribution of $Y_1$ given $Y_2$ is given as \eqref{mvb:BBconditionalpdf}.
\end{pf*}
%
%
\begin{pf*}{Proof of Lemma~\ref{mvb:biind}}
Expand the log-linear formulation of the bivariate Bernoulli
distribution~\eqref{mvb:BBloglinear} into factors
%
\begin{equation}
\label{proof:biind}P(Y_1 = y_1, Y_2 =
y_2) = p_{00} \exp\bigl(y_1f^1
\bigr) \exp \bigl(y_2f^2\bigr) \exp\bigl(y_1y_2f^{12}
\bigr).
\end{equation}
It is not hard to see that when $f^{12} = 0$, the density function
\eqref{proof:biind} is separable to two components with only $y_1$ and
$y_2$ in them. Therefore, the two random variables corresponding to the
formula are independent. Conversely, when $Y_1$ and $Y_2$ are
independent, their density function should be separable in terms of
$y_1$ and $y_2$, which implies $y_1y_2f^{12} = 0$ for any possible
values of $y_1$ and $y_2$. The assertion dictates that $f^{12}$ is zero.
\end{pf*}
%
%
\begin{pf*}{Proof of Lemma~\ref{mvb:transform}}
Consider the log-linear formulation \eqref{mvb:MBloglinear}, the
natural parameters $f$'s are combined with products of some components
of $y$. Let us match terms in the $f^{j_1\cdots j_r}B^{j_1\cdots
j_r}(y)$ from log-linear formulation \eqref{mvb:MBloglinear} with the
coefficient for the corresponding product $y_{j_1}\cdots y_{j_r}$ terms
in \eqref{mvb:pdf}. The exponents of $p$'s in \eqref{mvb:pdf} can be
expanded to summations of different products $B^\tau(y)$ with $\tau\in
\mathcal{T}$ and all the $p$'s with $y_{j_1},\ldots, y_{j_r}$ in the
exponent have effect on $f^{j_1\cdots j_r}$ so all the positions other
than $j_1,\ldots, j_r$ must be zero. Furthermore, those $p$'s with
positive $y_{j_1}\cdots y_{j_r}$ in its exponent appear in the
numerator of $\exp[f^{j_1\cdots j_r}]$ and the product is positive only
if there are even number of $0$'s in the positions $j_1, \ldots, j_r$.
The same scenario applies to the $p$'s with negative products in the exponents.

What's more, notice that $p_{00\cdots0} = b(\mathbf{f})$ and
%
\begin{eqnarray}
\exp\bigl[S^{j_1\cdots j_r}\bigr] &=& \exp\biggl[\sum
_{1\le s\le r} f^{j_s} + \sum_{1\le s<t\le r}
f^{j_sj_t} + \cdots+ f^{j_1j_2\cdots j_r}\biggr]
\nonumber
\\[-8pt]
\\[-8pt]
\nonumber
&=& \prod_{1\le s\le r}\exp\bigl[f^{j_s}\bigr]
\prod_{1\le s<t\le r} \exp \bigl[f^{j_sj_t}\bigr]\cdots
\exp\bigl[f^{j_1j_2\cdots j_r}\bigr]
\end{eqnarray}
and apply the formula for $\exp[f^{j_1\cdots j_r}]$ with cancellation
of terms in the numerators and the denominators. The resulting \eqref
{mvb:S} can then be verified.

Finally, \eqref{mvb:p} is a trivial extension of \eqref{mvb:S} by
exchanging the numerator and the denominator.
\end{pf*}
%
%
\begin{pf*}{Proof of Theorem~\ref{mvb:independence}}
Here, we take use of the moment generating function \eqref{mvb:mgf} but
it is also possible to directly work on the probability density
function \eqref{mvb:pdf}. The mgf can be rewritten as
%
\begin{equation}
\label{mvb:mgf2} \psi(\mu_1,\ldots, \mu_K) =
\frac{1}{\exp[b(\mathbf{f})]}\sum_{r=1}^K\sum
_{j_1\le j_2\le\cdots\le j_r} \exp\bigl[S^{j_1j_2\cdots
j_r}\bigr]\prod
_{k=1}^r\exp [\mu_{j_k} ].
\end{equation}
It is not hard to see that this is a polynomial function of the unknown
variables $\exp(\mu_k)$ for $k=1,\ldots, K$. The independence of the
random variables $Y_1, Y_2, \ldots, Y_K$ is equivalent to that~\eqref
{mvb:mgf2} can be separated into components of $\mu_k$ or equivalently
$\exp(\mu_k)$.

$(\Rightarrow)$ If the random vector $Y$ is independent, the moment
generating function should be separable and assume the formulation is
%
\begin{equation}
\label{mvb:mgf3} \psi(\mu_1, \ldots, \mu_K) = C\prod
_{k = 1}^K \bigl(\alpha_k +
\beta_k \exp [\mu_k]\bigr),
\end{equation}
where $\alpha_k$ and $\beta_k$ are functions of parameters $S$'s and
$C$ is a constant. If we expand \eqref{mvb:mgf3} to polynomial function
of $\exp[\mu_k]$ and determine the corresponding coefficients, \eqref
{mvb:independence1} and \eqref{mvb:independence2} will be derived.

$(\Leftarrow)$ Suppose \eqref{mvb:independence2} holds, then we have
\begin{eqnarray*}
\exp\bigl[S^{j_1j_2\cdots j_r}\bigr] = \prod_{k=1}^r
\exp\bigl[f^{j_k}\bigr],
\end{eqnarray*}
and as a result, the moment generating function can be decomposed to a
product of components of $\exp[\mu_k]$ like \eqref{mvb:mgf3} with the
following relations
\begin{eqnarray*}
C &=& \frac{1}{\exp[b(\mathbf{f})]},
\\
\alpha_k &=& 1,
\\
\beta_k &=& \exp\bigl[f^k\bigr].
\end{eqnarray*}
\upqed\end{pf*}
%

\begin{pf*}{Proof of Theorem~\ref{mvb:independence_group}}
The idea of proving the group independence of multivariate Bernoulli
variables are similar to Theorem~\ref{mvb:independence}. Instead of
decomposing the moment generating function to products of $\mu_k$, we
only have to separate them into groups with each only involving the
dependent random variables. That is to say, the moment generating
function with two separately independent nodes in the multivariate
Bernoulli should have the form
\begin{eqnarray*}
&&\psi(\mu_1, \ldots, \mu_K)\\
&&\qquad = \bigl(\alpha_0
+ \alpha_1\exp[\mu_1] + \cdots + \alpha_r
\exp[\mu_r]\bigr)\cdot\bigl(\beta_0 + \beta_1
\exp[\mu_{r+1}] + \cdots+ \beta_s\exp[\mu_{K}]
\bigr).
\end{eqnarray*}
Matching the corresponding coefficients of this separable moment
generating function and the natural parameters leads to the conclusion
\eqref{mvb:ind_group}.
\end{pf*}
\end{appendix}
%
%

\section*{Acknowledgements}
Research of all three authors was supported in part by NIH Grant EY09946
and NSF Grant DMS-09-06818.

%

%



\printhistory


\begin{thebibliography}{25}

\bibitem{Banerjee:2008}
\begin{barticle}[mr]
\bauthor{\bsnm{Banerjee},~\bfnm{Onureena}\binits{O.}},
\bauthor{\bsnm{El~Ghaoui},~\bfnm{Laurent}\binits{L.}} \AND
\bauthor{\bsnm{d'Aspremont},~\bfnm{Alexandre}\binits{A.}}
(\byear{2008}).
\btitle{Model selection through sparse maximum likelihood estimation for
multivariate {G}aussian or binary data}.
\bjournal{J. Mach. Learn. Res.}
\bvolume{9}
\bpages{485--516}.
\bid{issn={1532-4435}, mr={2417243}}
\bptok{imsref}%
\end{barticle}
\endbibitem

\bibitem{Craven:1978}
\begin{barticle}[mr]
\bauthor{\bsnm{Craven},~\bfnm{Peter}\binits{P.}} \AND
\bauthor{\bsnm{Wahba},~\bfnm{Grace}\binits{G.}}
(\byear{1979}).
\btitle{Smoothing noisy data with spline functions. {E}stimating the correct
degree of smoothing by the method of generalized cross-validation}.
\bjournal{Numer. Math.}
\bvolume{31}
\bpages{377--403}.
\bid{doi={10.1007/BF01404567}, issn={0029-599X}, mr={0516581}}
\bptnote{check year}%
\bptok{imsref}%
\end{barticle}
\endbibitem

\bibitem{Dai:2012}
\begin{bmisc}[author]
\bauthor{\bsnm{Dai},~\bfnm{Bin}\binits{B.}}
(\byear{2012}).
\bhowpublished{Multivariate Bernoulli distribution models. Technical report.
Dept. Statistics, Univ. Wisconsin, Madison, WI 53706.}
\bptok{imsref}%
\end{bmisc}
\endbibitem

\bibitem{Ding:2011}
\begin{bincollection}[author]
\bauthor{\bsnm{Ding},~\bfnm{S.}\binits{S.}},
\bauthor{\bsnm{Wahba},~\bfnm{G.}\binits{G.}} \AND
\bauthor{\bsnm{Zhu},~\bfnm{X.}\binits{X.}}
(\byear{2011}).
\btitle{Learning higher-order graph structure with features by structure
penalty}.
In \bbooktitle{Advances in Neural Information Processing Systems}
\bvolume{24}
\bpages{253--261}.
\bnote{25th Annual Conference on Neural Information Processing Systems 2011.
Proceedings of a meeting held 12--14 December 2011, Granada, Spain}.
\bptok{imsref}%
\end{bincollection}
\endbibitem

\bibitem{Friedman:2010}
\begin{barticle}[author]
\bauthor{\bsnm{Friedman},~\bfnm{J.}\binits{J.}},
\bauthor{\bsnm{Hastie},~\bfnm{T.}\binits{T.}} \AND
\bauthor{\bsnm{Tibshirani},~\bfnm{R.}\binits{R.}}
(\byear{2010}).
\btitle{Regularization paths for generalized linear models via coordinate
descent}.
\bjournal{Journal of Statistical Software}
\bvolume{44}
\bpages{1--22}.
\bptok{imsref}%
\end{barticle}
\endbibitem

\bibitem{Gao:2001}
\begin{barticle}[mr]
\bauthor{\bsnm{Gao},~\bfnm{Fangyu}\binits{F.}},
\bauthor{\bsnm{Wahba},~\bfnm{Grace}\binits{G.}},
\bauthor{\bsnm{Klein},~\bfnm{Ronald}\binits{R.}} \AND
\bauthor{\bsnm{Klein},~\bfnm{Barbara}\binits{B.}}
(\byear{2001}).
\btitle{Smoothing spline {ANOVA} for multivariate {B}ernoulli observations,
with application to ophthalmology data}.
\bjournal{J. Amer. Statist. Assoc.}
\bvolume{96}
\bpages{127--160}.
\bid{doi={10.1198/016214501750332749}, issn={0162-1459}, mr={1952725}}
\bptnote{check related}%
\bptok{imsref}%
\end{barticle}
\endbibitem

\bibitem{Gu:2002}
\begin{bbook}[mr]
\bauthor{\bsnm{Gu},~\bfnm{Chong}\binits{C.}}
(\byear{2002}).
\btitle{Smoothing Spline {ANOVA} Models}.
\bseries{Springer Series in Statistics}.
\blocation{New York}: \bpublisher{Springer}.
\bid{mr={1876599}}
\bptok{imsref}%
\end{bbook}
\endbibitem

\bibitem{Ising:1925}
\begin{barticle}[author]
\bauthor{\bsnm{Ising},~\bfnm{E.}\binits{E.}}
(\byear{1925}).
\btitle{Beitrag zur Theorie des Ferromagnetismus}.
\bjournal{Z. Phys.}
\bvolume{31}
\bpages{253--258}.
\bptok{imsref}%
\end{barticle}
\endbibitem

\bibitem{Ma:2010}
\begin{bmisc}[author]
\bauthor{\bsnm{Ma},~\bfnm{X.}\binits{X.}}
(\byear{2010}).
\bhowpublished{Penalized regression in reproducing kernel Hilbert spaces with
randomized covariate data. Technical Report No. 1159. Dept. Statistics, Univ. Wisconsin, Madison, WI 53706.}
\bptok{imsref}%
\end{bmisc}
\endbibitem

\bibitem{McCullagh:1989}
\begin{bbook}[author]
\bauthor{\bsnm{McCullagh},~\bfnm{P.}\binits{P.}} \AND
\bauthor{\bsnm{Nelder},~\bfnm{J.}\binits{J.}}
(\byear{1989}).
\btitle{Generalized Linear Models}.
\blocation{New York}: \bpublisher{Chapman \& Hall}.
\bptok{imsref}%
\end{bbook}
\endbibitem

\bibitem{Meinshausen:2006}
\begin{barticle}[mr]
\bauthor{\bsnm{Meinshausen},~\bfnm{Nicolai}\binits{N.}} \AND
\bauthor{\bsnm{B{\"u}hlmann},~\bfnm{Peter}\binits{P.}}
(\byear{2006}).
\btitle{High-dimensional graphs and variable selection with the lasso}.
\bjournal{Ann. Statist.}
\bvolume{34}
\bpages{1436--1462}.
\bid{doi={10.1214/009053606000000281}, issn={0090-5364}, mr={2278363}}
\bptok{imsref}%
\end{barticle}
\endbibitem

\bibitem{Park:2008}
\begin{barticle}[mr]
\bauthor{\bsnm{Park},~\bfnm{Trevor}\binits{T.}} \AND
\bauthor{\bsnm{Casella},~\bfnm{George}\binits{G.}}
(\byear{2008}).
\btitle{The {B}ayesian lasso}.
\bjournal{J. Amer. Statist. Assoc.}
\bvolume{103}
\bpages{681--686}.
\bid{doi={10.1198/016214508000000337}, issn={0162-1459}, mr={2524001}}
\bptok{imsref}%
\end{barticle}
\endbibitem

\bibitem{Ravikumar:2010}
\begin{barticle}[mr]
\bauthor{\bsnm{Ravikumar},~\bfnm{Pradeep}\binits{P.}},
\bauthor{\bsnm{Wainwright},~\bfnm{Martin~J.}\binits{M.J.}} \AND
\bauthor{\bsnm{Lafferty},~\bfnm{John~D.}\binits{J.D.}}
(\byear{2010}).
\btitle{High-dimensional {I}sing model selection using {$\ell_1$}-regularized logistic regression}.
\bjournal{Ann. Statist.}
\bvolume{38}
\bpages{1287--1319}.
\bid{doi={10.1214/09-AOS691}, issn={0090-5364}, mr={2662343}}
\bptok{imsref}%
\end{barticle}
\endbibitem

\bibitem{Shi:2012}
\begin{barticle}[author]
\bauthor{\bsnm{Shi},~\bfnm{Weiliang}\binits{W.}},
\bauthor{\bsnm{Wahba},~\bfnm{Grace}\binits{G.}},
\bauthor{\bsnm{Irizarry},~\bfnm{R.}\binits{R.}},
\bauthor{\bsnm{Corrado~Bravo},~\bfnm{Hector}\binits{H.}} \AND
\bauthor{\bsnm{Wright},~\bfnm{Stephen}\binits{S.}}
(\byear{2012}).
\btitle{The partitioned LASSO-patternsearch algorithm with application to gene
expression data}.
\bjournal{BMC Bioinformatics}
\bvolume{13}
\bpages{98--110}.
\bptok{imsref}%
\end{barticle}
\endbibitem

\bibitem{Shi:2008}
\begin{barticle}[mr]
\bauthor{\bsnm{Shi},~\bfnm{Weiliang}\binits{W.}},
\bauthor{\bsnm{Wahba},~\bfnm{Grace}\binits{G.}},
\bauthor{\bsnm{Wright},~\bfnm{Stephen}\binits{S.}},
\bauthor{\bsnm{Lee},~\bfnm{Kristine}\binits{K.}},
\bauthor{\bsnm{Klein},~\bfnm{Ronald}\binits{R.}} \AND
\bauthor{\bsnm{Klein},~\bfnm{Barbara}\binits{B.}}
(\byear{2008}).
\btitle{L{ASSO}-{P}atternsearch algorithm with application to ophthalmology and
genomic data}.
\bjournal{Stat. Interface}
\bvolume{1}
\bpages{137--153}.
\bid{issn={1938-7989}, mr={2425351}}
\bptok{imsref}%
\end{barticle}
\endbibitem

\bibitem{Tibshirani:1996}
\begin{barticle}[mr]
\bauthor{\bsnm{Tibshirani},~\bfnm{Robert}\binits{R.}}
(\byear{1996}).
\btitle{Regression shrinkage and selection via the lasso}.
\bjournal{J. R. Stat. Soc. Ser. B Stat. Methodol.}
\bvolume{58}
\bpages{267--288}.
\bid{issn={0035-9246}, mr={1379242}}
\bptok{imsref}%
\end{barticle}
\endbibitem

\bibitem{Wahba:1990}
\begin{bbook}[mr]
\bauthor{\bsnm{Wahba},~\bfnm{Grace}\binits{G.}}
(\byear{1990}).
\btitle{Spline Models for Observational Data}.
\bseries{CBMS-NSF Regional Conference Series in Applied Mathematics}
\bvolume{59}.
\blocation{Philadelphia, PA}: \bpublisher{SIAM}.
\bid{doi={10.1137/1.9781611970128}, mr={1045442}}
\bptok{imsref}%
\end{bbook}
\endbibitem

\bibitem{Wahba:1995}
\begin{barticle}[mr]
\bauthor{\bsnm{Wahba},~\bfnm{Grace}\binits{G.}},
\bauthor{\bsnm{Wang},~\bfnm{Yuedong}\binits{Y.}},
\bauthor{\bsnm{Gu},~\bfnm{Chong}\binits{C.}},
\bauthor{\bsnm{Klein},~\bfnm{Ronald}\binits{R.}} \AND
\bauthor{\bsnm{Klein},~\bfnm{Barbara}\binits{B.}}
(\byear{1995}).
\btitle{Smoothing spline {ANOVA} for exponential families, with application to
the {W}isconsin {E}pidemiological {S}tudy of {D}iabetic {R}etinopathy}.
\bjournal{Ann. Statist.}
\bvolume{23}
\bpages{1865--1895}.
\bid{doi={10.1214/aos/1034713638}, issn={0090-5364}, mr={1389856}}
\bptok{imsref}%
\end{barticle}
\endbibitem

\bibitem{Wainwright:2008}
\begin{barticle}[author]
\bauthor{\bsnm{Wainwright},~\bfnm{M.}\binits{M.}} \AND
\bauthor{\bsnm{Jordan},~\bfnm{M.}\binits{M.}}
(\byear{2008}).
\btitle{Graphical models, exponential families, and variational inference}.
\bjournal{Foundations and Trends in Machine Learning}
\bvolume{1}
\bpages{1--305}.
\bptok{imsref}%
\end{barticle}
\endbibitem

\bibitem{Whittaker:1990}
\begin{bbook}[author]
\bauthor{\bsnm{Whittaker},~\bfnm{J.}\binits{J.}}
(\byear{1990}).
\btitle{Graphical Models in Applied Mathematical Multivariate Statistics}.
\blocation{New York}: \bpublisher{Wiley}.
\bptok{imsref}%
\end{bbook}
\endbibitem

\bibitem{Xiang:1994}
\begin{bmisc}[author]
\bauthor{\bsnm{Xiang},~\bfnm{D.}\binits{D.}} \AND
\bauthor{\bsnm{Wahba},~\bfnm{G.}\binits{G.}}
(\byear{1994}).
\bhowpublished{A generalized approximate cross validation for smoothing splines with
non-Gaussian data. Technical Report No. 930. Dept. Statistics, Univ. Wisconsin, Madison, WI 53706.}
\bptok{imsref}%
\end{bmisc}
\endbibitem

\bibitem{Xue:2012}
\begin{barticle}[mr]
\bauthor{\bsnm{Xue},~\bfnm{Lingzhou}\binits{L.}},
\bauthor{\bsnm{Zou},~\bfnm{Hui}\binits{H.}} \AND
\bauthor{\bsnm{Cai},~\bfnm{Tianxi}\binits{T.}}
(\byear{2012}).
\btitle{Nonconcave penalized composite conditional likelihood estimation of
sparse {I}sing models}.
\bjournal{Ann. Statist.}
\bvolume{40}
\bpages{1403--1429}.
\bid{doi={10.1214/12-AOS1017}, issn={0090-5364}, mr={3015030}}
\bptok{imsref}%
\end{barticle}
\endbibitem

\bibitem{Zhao:2006}
\begin{barticle}[mr]
\bauthor{\bsnm{Zhao},~\bfnm{Peng}\binits{P.}} \AND
\bauthor{\bsnm{Yu},~\bfnm{Bin}\binits{B.}}
(\byear{2006}).
\btitle{On model selection consistency of {L}asso}.
\bjournal{J. Mach. Learn. Res.}
\bvolume{7}
\bpages{2541--2563}.
\bid{issn={1532-4435}, mr={2274449}}
\bptok{imsref}%
\end{barticle}
\endbibitem

\bibitem{Zhao:2007}
\begin{barticle}[mr]
\bauthor{\bsnm{Zhao},~\bfnm{Peng}\binits{P.}} \AND
\bauthor{\bsnm{Yu},~\bfnm{Bin}\binits{B.}}
(\byear{2007}).
\btitle{Stagewise lasso}.
\bjournal{J. Mach. Learn. Res.}
\bvolume{8}
\bpages{2701--2726}.
\bid{issn={1532-4435}, mr={2383572}}
\bptok{imsref}%
\end{barticle}
\endbibitem

\end{thebibliography}
\end{document}